# Transceiver Design for GFDM with Index Modulation in Multi-user Networks

Merve Yüzgeçcioğlu and Eduard Jorswieck
Communications Theory, Communications Laboratory
Dresden University of Technology, D-01062 Dresden, Germany
Email: {merve.yuzgeccioglu, eduard.jorswieck}@tu-dresden.de

*Abstract*—Index modulation (IM) techniques can be applied to the different media in order to achieve spectral- and energy-efficient communication as well as to the indices of the subcarriers of a generalized frequency division multiplexing (GFDM) data block. In this work, a novel transceiver architecture for multi-user GFDM-IM system is introduced. The performance of the GFDM-IM is studied by considering the bit error rate (BER) as performance metric. It is shown that better BER performance than the classical GFDM and the orthogonal frequency division multiplexing (OFDM) with IM can be achieved by employing IM to the GFDM.

## I. INTRODUCTION

Multicarrier waveforms are widely used due to their high spectral efficiency natures. Orthogonal frequency division multiplexing (OFDM) multicarrier communication scheme is already implemented for several systems including 4G long-term evolution (LTE). However, the demand on spectral efficiency for 5G wireless networks is much larger than the current 4G networks, since 5G networks will involve not only the cellular communications but also several different players such as sensors for smart cities, device to device communications, self driving vehicles, etc. Hence, there is a need for new modulation scheme that will provide higher spectrum efficiency with better energy efficiency.

Index modulation (IM) schemes have attracted significant attention that allow to transmit additional bits to the conventional modulation schemes by mapping the information to the indices of the different medias. OFDM-IM became popular recently that allows to map information to the indices of the subcarriers. This scheme firstly introduced in [1] as subcarrier index modulation (SIM) OFDM. At this work, number of active subcarriers varies according to the incoming bit stream. Later on, another method is introduced in [2], [3] where the number of active subcarriers are predefined and can be adjusted according to the system needs. There has been extensive amount of work on different IM methods for OFDM in order to achieve better spectral efficiency, better error performance, etc. [4]–[12]. Furthermore, a survey on IM schemes is recently published in [13] that provides a broader look into different applications and performance evaluations.

Generalized frequency division multiplexing (GFDM) is another multicarrier transmission scheme that is introduced to literature in [14]. A low complexity matrix model that provides a practical method in order to generate the GFDM block is introduced in [16] and the performance of the GFDM scheme is compared with the LTE standard in spectral properties and implementation complexity aspects. In [15], matrix model is further improved and linear receiver structures are introduced. In addition, at this work the bit error rate (BER) performance is studied. In [17], possible applications of GFDM scheme for different 5G network scenarios are further studied.

Furthermore, IM technique is combined with the GFDM in [18] and it is shown that GFDM-IM outperforms the error performance of GFDM at mid- to high-SNR values. Combination of spatial modulation (SM) that is mapping the information to the indices of the transmit antennas and GFDM subcarriers is studied in [19]. In addition, space and frequency index modulation (SFIM) scheme is introduced for MIMO GFDM systems in [20]. In this work, active transmit antenna index, subcarrier index and the constellation symbols are determined according to the incoming bit stream. It is shown that the SFIM-GFDM outperforms the SM-GFDM.

In this work, transceiver design of the GFDM-IM system in multi-user networks is introduced. The transmitter structure is explained in detail in order to construct the GFDM-IM symbol at user side. Moreover, the receiver structure is explained in order to mitigate the inter-user-interference (IUI) and successfully decode the active subcarrier indices and the conventionally modulated data. Error performance of the system is studied and it is shown that the GFDM-IM scheme achieves better error performance than both the OFDM-IM and the classical GFDM scheme.

The rest of the paper is organized as follows. In Sec. II, system model of multi-user GFDM-IM is explained in detail. Performance of the system is studied in Sec. III and the paper is concluded in Sec. IV.

## II. SYSTEM MODEL

In a classical GFDM scheme, there are $L$ subsymbols each contains $N_{tot}$ available subcarriers. During the transmission of a GFDM block, total $Q = LN_{tot}$ symbols are transmitted that allows us to transmit $Q \log_2 M$ bits where $M$ is the modulation order. In contrary, not all the available subcarriers at a subsymbol actively carry information on the GFDM-IM scheme. According to the incoming bit stream, $K_{tot}$ out of $N_{tot}$ subcarriers are chosen as active and carry

The work of Merve Yüzgeçcioğlu has received partly funding from the European Union's Horizon 2020 research and innovation programme under the Marie Sklodowzka-Curie grant agreement No 641985.

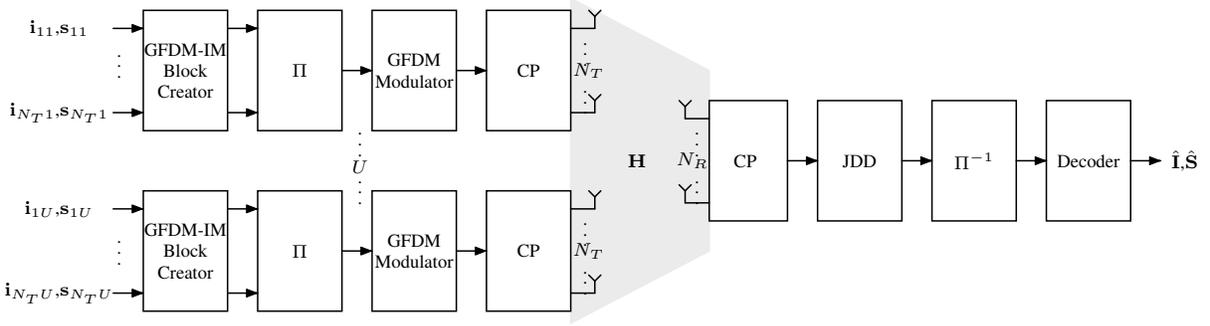

Fig. 1. Block diagram of GFDM-IM uplink scenario in multi-user networks

$M$-ary modulated data at each subsymbol. The remaining $L(N_{tot} - K_{tot})$ subcarriers remain idle and the location of the active subcarriers carries additional information to the conventionally modulated data.

The considered uplink system model is depicted in Fig. 1. At this system model, there are $U$ users with $N_T$ transmit antennas. The base station (BS) is equipped with $N_R$ receive antennas where $N_R \geq UN_T$. It is assumed that BS has CSI and users are not aware of the channel statistics. At each transmit antenna chain of a user, incoming bit stream is firstly splitted into $L$ groups for each subsymbol. Then, the available $N_{tot}$ subcarriers of each subsymbol is divided into $G$ groups where $N = N_{tot}/G$. Furthermore, according to the bits in one of the groups, $K$ subcarriers are selected active by the help of a predefined look-up table where $K = K_{tot}/G$. An example of the look-up table $\boldsymbol{\Phi}$ is given for $N = 4$ and $K = 2$

$$\boldsymbol{\Phi} = \begin{bmatrix} 1 & 1 & 1 & 2 \\ 2 & 3 & 4 & 3 \end{bmatrix}^T. \tag{1}$$

Note that, according to the number of bits represented by IM part of the system, the look-up table is truncated. The resulting total number of bits transmitted from one of the transmit antennas with a GFDM-IM block is $b = LG(b_1 + b_2)$ where $b_1 = \lfloor \log_2 \binom{N}{K} \rfloor$ by IM part and $b_2 = K \log_2 M$ as $M$-ary modulated symbols. The indices of the selected subcarriers are stored in the vector $\mathbf{i}_{ltu}^g = [i_{ltu}^g(1), i_{ltu}^g(2), \ldots, i_{ltu}^g(K)]^T$ where $g = 1, \ldots, G$, $l = 1, \ldots, L$, $t = 1, \ldots, N_T$ and $u = 1, \ldots, U$. Here, $i_{ltu}^g(k) = 1, \ldots, N$ is the selected subcarrier index of the $g$-th group on the $l$-th subsymbol at the $u$-th user to transmit from the $t$-th transmit antenna. Furthermore, the $M$-ary modulated symbols are stored in the vector $\mathbf{s}_{ltu}^g = [s_{ltu}^g(1), s_{ltu}^g(2), \ldots, s_{ltu}^g(K)]^T$ to transmit on the selected subcarriers. The resulting subcarrier indices and the modulated symbols for all groups are $\mathbf{i}_{ltu} = [(\mathbf{i}_{ltu}^1)^T, (\mathbf{i}_{ltu}^2)^T, \ldots, (\mathbf{i}_{ltu}^G)^T]^T$ and $\mathbf{s}_{ltu} = [(\mathbf{s}_{ltu}^1)^T, (\mathbf{s}_{ltu}^2)^T, \ldots, (\mathbf{s}_{ltu}^G)^T]^T$, respectively, where $\mathbf{i}_{ltu}$ and $\mathbf{s}_{ltu}$ are $\in \mathbb{C}^{K_{tot} \times 1}$.

Furthermore, the modulated data are assigned to the active subcarriers and the data vector of the $l$-th subsymbol and the $g$-th group $\mathbf{d}_{ltu}^g = [d_{ltu}^g(1), d_{ltu}^g(2), \ldots, d_{ltu}^g(N)]^T$ is generated. Note that $N - K$ subcarriers on the vector $\mathbf{d}_{ltu}^g$ remain idle during transmission. Then, the data vector of each subsymbol is generated by combining all the gorups $\mathbf{d}_{ltu} = [(\mathbf{d}_{ltu}^1)^T, (\mathbf{d}_{ltu}^2)^T, \ldots, (\mathbf{d}_{ltu}^G)^T]^T$. The data block is generated by combining the data of all subsymbols $\mathbf{d}_{tu} = [\mathbf{d}_{1tu}^T, \mathbf{d}_{2tu}^T \ldots, \mathbf{d}_{Ltu}^T]^T$. Finally, a block interleaver is employed in order to transmit each subcarrier in a group through uncorrelated channels and the GFDM-IM data block of the $u$-th user to transmit from the $t$-th transmit antenna $\bar{\mathbf{d}}_{tu} \in \mathbb{C}^{Q \times 1}$ is generated.

After generating the GFDM-IM data block, the remaining procedure is same as the classical GFDM scheme. Moreover, the resulting GFDM symbol can be generated as follows

$$\mathbf{x}_{tu} = \mathbf{A}\bar{\mathbf{d}}_{tu}. \tag{2}$$

Herein, $\mathbf{A}$ is the $Q \times Q$ transmitter matrix generated as given in [15]. $\mathbf{A}$ has the structure of

$$\mathbf{A} = [\mathbf{g}_{0,0}, \ldots, \mathbf{g}_{N_{tot}-1,0}, \mathbf{g}_{0,1}, \ldots, \mathbf{g}_{N_{tot}-1,1}, \mathbf{g}_{0,L-1}, \\ \ldots, \mathbf{g}_{N_{tot}-1,L-1}] \tag{3}$$

$$\text{where} \quad g_{n,l} = g((q-nL)_{\text{mod} Q}) \exp j2\pi \frac{nq}{N_{tot}} \tag{4}$$

is the circularly shifted transmit filter and $\mathbf{g}_{n,l} = [g_{n,l}(0), g_{n,l}(1), \ldots g_{n,l}(Q-1)]^T$. The transmitted data from the $u$-th user can be written as $\mathbf{x}_u = [\mathbf{x}_{1u}^T, \mathbf{x}_{2u}^T, \ldots, \mathbf{x}_{Tu}^T]^T$. Finally, the received symbol at the BS is

$$\mathbf{y} = \sum_{u=1}^{U} \mathbf{H}_u \mathbf{x}_u + \mathbf{w}, \tag{5}$$

where $\mathbf{w}_n \in \mathbb{C}^{QN_R \times 1}$ is the AWGN with $\mathcal{CN}(0, \sigma^2)$ distribution. $\mathbf{H}_u$ has the following structure

$$\mathbf{H}_u = \begin{bmatrix} \mathbf{H}_{11u} & \mathbf{H}_{12u} & \ldots & \mathbf{H}_{1Tu} \\ \vdots & \ddots & \vdots \\ \mathbf{H}_{R1u} & \mathbf{H}_{R2u} & \ldots & \mathbf{H}_{RTu} \end{bmatrix}, \tag{6}$$

where $\mathbf{H}_{rtu} \in \mathbb{C}^{Q \times Q}$ is the circular convolution matrix generated from the channel impulse response coefficients $\mathbf{h}_{tu} = [h_{tu1}, h_{tu2}, \ldots, h_{tuV}]$ where $V$ is the number of channel taps and $h_{tuv}$ are circularly symmetric complex Gaussian random variables with $\mathcal{CN}(0, 1/V)$.

We employ MMSE filtering in order to jointly perform the detection and the GFDM demodulation (JDD) [20]. The MMSE filter is $\mathbf{W} = (\mathbf{B}^H\mathbf{B} + \sigma^2\mathbf{I})^{-1}\mathbf{B}^H$ where $\mathbf{B} = [\mathbf{B}_1, \mathbf{B}_2, \ldots, \mathbf{B}_U]$ with

$$\mathbf{B}_u = \begin{bmatrix} \mathbf{H}_{11u}\mathbf{A} & \mathbf{H}_{12u}\mathbf{A} & \ldots & \mathbf{H}_{1Tu}\mathbf{A} \\ & \vdots & \ddots & \vdots \\ \mathbf{H}_{R1u}\mathbf{A} & \mathbf{H}_{R2u}\mathbf{A} & \ldots & \mathbf{H}_{RTu}\mathbf{A} \end{bmatrix}. \quad (7)$$

The resulting estimated vector $\tilde{\mathbf{d}} = \mathbf{W}\mathbf{y} \in \mathbb{C}^{QN_TU\times 1}$ contains the transmitted data from all the users. Furthermore, the estimated data vector of each user is divided into $L$ groups for each subsymbol $\hat{\mathbf{d}}_{tu} = [\hat{\mathbf{d}}_{1tu}^T, \hat{\mathbf{d}}_{2tu}^T \ldots, \hat{\mathbf{d}}_{Ltu}^T]^T$. Finally, each estimated subsymbol is divided into $G$ groups in order to decode the index modulated and $M$-ary modulated bits $\hat{\mathbf{d}}_{ltu} = [(\hat{\mathbf{d}}_{ltu}^1)^T, (\hat{\mathbf{d}}_{ltu}^2)^T, \ldots, (\hat{\mathbf{d}}_{ltu}^G)^T]^T$ where $l = 1, \ldots, L$, $t = 1, \ldots, N_T$ and $u = 1, \ldots, U$.

In order to detect the active subcarriers in the received data $\hat{\mathbf{d}}_{ltu}^g \in \mathbb{C}^{N\times 1}$, we calculate the decision metric for each possible combination of the active subcarriers by the help of the look-up table $\mathbf{\Phi}$

$$\mathbf{m}(j) = \sum_{k=1}^{K} |\hat{\mathbf{d}}_{ltu}^g(\mathbf{\Phi}(j,k))|, \quad (8)$$

where $j = 1, \ldots, 2^{b_1}$, $g = 1, \ldots, G$, $l = 1, \ldots, L$, $t = 1, \ldots, N_T$, and $u = 1, \ldots, U$. Once we have the decision metric $\mathbf{m}$, the maximum entry of this metric that also indicates the active subcarrier combination is found as $\hat{l} = \arg\max_l \mathbf{m}(l)$. Finally, $\hat{l}$ is mapped to $\mathbf{\Phi}$ and the active subcarriers are detected as follows

$$\hat{\mathbf{i}}_{ltu}^g = \mathbf{\Phi}(\hat{l}, :). \quad (9)$$

Once the active subcarriers are detected, the transmitted symbols on these subcarriers can be easily demodulated by classical $M$-ary demodulation. Finally, $\hat{\mathbf{i}}_{ltu}^g \in \mathbb{C}^{K\times 1}$ contains the detected active subcarrier indices and $\hat{\mathbf{s}}_{ltu}^g \in \mathbb{C}^{K\times 1}$ contains the detected $M$-ary modulated symbols of the $g$-th group of the $l$-th subsymbol transmitted from the $t$-th antenna of the $u$-th user. The resulting detected indices and $M$-ary modulated data of all the users are $[\hat{\mathbf{I}}_1, \ldots, \hat{\mathbf{I}}_U]$ and $[\hat{\mathbf{S}}_1, \ldots, \hat{\mathbf{S}}_U]$, respectively, where $\hat{\mathbf{I}}_u$ and $\hat{\mathbf{S}}_u$ are $\in \mathbb{C}^{LK_{tot}\times N_T}$ for $u = 1, \ldots, U$.

## III. NUMERICAL RESULTS

In this section, the BER performance of the GFDM-IM is compared to the GFDM and the OFDM-IM schemes. In the simulation setup, the channel is considered i.i.d. multi-path Rayleigh fading. 2- and 4-user networks are considered in order to study the performance of the schemes and the BS is equipped with the same number of receive antennas as the number of users. At one transmission, $L = 5$ subsymbols and $N_{tot} = 128$ subcarriers are used. Number of available subcarriers in a group is determined as $N = 4$ and $K = 2$ of them are chosen at each transmission. Additionally, 4-QAM is considered to modulate the active subcarriers. Data is filtered with square-root-raise-cosine filter with roll-off factors $\alpha = 0.1$ and $\alpha = 0.9$ to generate the GFDM block.

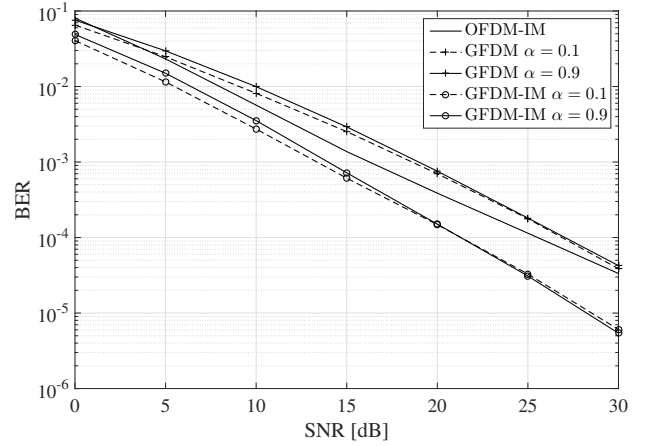

Fig. 2. Bit error rate comparison of OFDM-IM, GFDM and GFDM-IM schemes in 2-user network ($N_T = 1$, $N_R = 2$, $L = 5$, $N_{tot} = 128$, $N = 4$, $K = 2$ with 4-QAM)

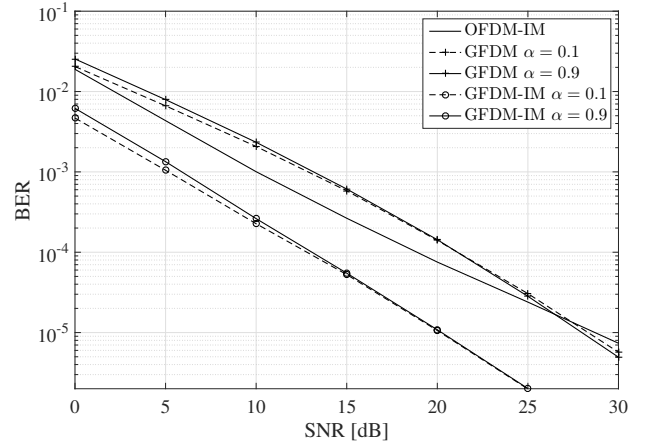

Fig. 3. Bit error rate comparison of OFDM-IM, GFDM and GFDM-IM schemes in 4-user network ($N_T = 1$, $N_R = 4$, $L = 5$, $N_{tot} = 128$, $N = 4$, $K = 2$ with 4-QAM)

In Fig. 2, the performance of GFDM-IM is compared to the classical GFDM and the OFDM-IM schemes for 2-user network. It is observed that the performance of the GFDM-IM schemes is better than both the classical GFDM and the OFDM-IM schemes. This performance improvement is the result of the MMSE-based JDD and the improved distance of the symbols in frequency domain caused by the IM. Also note that the increasing the roll-off factor results a slightly worse performance up to mid-SNR values.

Performance of the system for higher number of users is examined in Fig. 3. We can see in this simulation result that the behavior of the three schemes are similar to the 2-user case. The BER performance of the all three system gets better due to the increased number of receive antennas at BS side.

## IV. Conclusion

In this work, the GFDM-IM scheme has been studied for multi-user networks. The receiver architecture was introduced in order to decode the received data successfully when interference from different users also presents. It has been shown that when the IM was employed to the subcarriers of a GFDM block, error performance of the system improves. Moreover, GFDM-IM achieves better performance than both the classical GFDM and OFDM-IM schemes. Furthermore, better spectral efficiency can be achieved by arranging the system parameters or the transmit power can be reduced by lowering the number of active subcarriers without loosing the spectral efficiency much. As a conclusion, GFDM-IM offers a flexible system model with performance improvement compared to the state-of-the-art schemes.

## References

[1] R. Abu-alhiga and H. Haas, "Subcarrier-index modulation OFDM," in *IEEE 20th International Symposium on Personal, Indoor and Mobile Radio Communications*, Sep. 2009, pp. 177–181.

[2] E. Başar, Ü. Aygölü, E. Panayırcı, and H. V. Poor, "Orthogonal frequency division multiplexing with index modulation," in *IEEE Global Communications Conference (GLOBECOM)*, Dec. 2012, pp. 4741–4746.

[3] ——, "Orthogonal frequency division multiplexing with index modulation," *IEEE Transactions on Signal Processing*, vol. 61, no. 22, pp. 5536–5549, Nov. 2013.

[4] E. Başar, "Multiple-input multiple-output OFDM with index modulation," *IEEE Signal Processing Letters*, vol. 22, no. 12, pp. 2259–2263, Dec. 2015.

[5] L. Gong, L. Dan, S. Feng, S. Wang, Y. Xiao, and S. Li, "Subcarrier-index based vector-modulated IFDMA systems," in *International Conference on Communications, Circuits and Systems (ICCCAS)*, vol. 2, Nov. 2013, pp. 19–21.

[6] M. Wen, X. Cheng, M. Ma, B. Jiao, and H. V. Poor, "On the achievable rate of OFDM with index modulation," *IEEE Transactions on Signal Processing*, vol. 64, no. 8, pp. 1919–1932, Apr. 2016.

[7] Y. Ko, "A tight upper bound on bit error rate of joint OFDM and multi-carrier index keying," *IEEE Communications Letters*, vol. 18, no. 10, pp. 1763–1766, Oct. 2014.

[8] Y. Xiao, S. Wang, L. Dan, X. Lei, P. Yang, and W. Xiang, "OFDM with interleaved subcarrier-index modulation," *IEEE Communications Letters*, vol. 18, no. 8, pp. 1447–1450, Aug. 2014.

[9] T. Datta, H. S. Eshwaraiah, and A. Chockalingam, "Generalized space-and-frequency index modulation," *IEEE Transactions on Vehicular Technology*, vol. 65, no. 7, pp. 4911–4924, Jul. 2016.

[10] B. Chakrapani, T. L. Narasimhan, and A. Chockalingam, "Generalized space-frequency index modulation: Low-complexity encoding and detection," in *IEEE Globecom Workshops*, Dec. 2015, pp. 1–6.

[11] H. Zhu, W. Wang, Q. Huang, and X. Gao, "Subcarrier index modulation OFDM for multiuser MIMO systems with iterative detection," in *IEEE International Symposium on Personal, Indoor, and Mobile Radio Communications*, Sep. 2016, pp. 1–6.

[12] ——, "Uplink transceiver for subcarrier index modulation OFDM in massive MIMO systems with imperfect channel state information," in *International Conference on Wireless Communications Signal Processing*, Oct. 2016, pp. 1–6.

[13] E. Başar, "Index modulation techniques for 5G wireless networks," *IEEE Communications Magazine*, vol. 54, no. 7, pp. 168–175, Jul. 2016.

[14] G. Fettweis, M. Krondorf, and S. Bittner, "GFDM - generalized frequency division multiplexing," in *VTC Spring 2009 - IEEE 69th Vehicular Technology Conference*, Apr. 2009, pp. 1–4.

[15] N. Michailow, S. Krone, M. Lentmaier, and G. Fettweis, "Bit error rate performance of generalized frequency division multiplexing," in *IEEE Vehicular Technology Conference*, Sep. 2012, pp. 1–5.

[16] N. Michailow, I. Gaspar, S. Krone, M. Lentmaier, and G. Fettweis, "Generalized frequency division multiplexing: Analysis of an alternative multi-carrier technique for next generation cellular systems," in *International Symposium on Wireless Communication Systems*, Aug. 2012, pp. 171–175.

[17] N. Michailow, M. Matthé, I. S. Gaspar, A. N. C. and Luciano Leonel Mendes, A. Festag, and G. Fettweis, "Generalized frequency division multiplexing for 5th generation cellular networks," *IEEE Transactions on Communications*, vol. 62, no. 9, pp. 3045–3061, Sep. 2014.

[18] E. Öztürk, E. Başar, and H. A. Çırpan, "Generalized frequency division multiplexing with index modulation," in *IEEE Global Communications Conference*, Dec. 2016, pp. 1–6.

[19] E. Öztürk, E. Basar, and H. A. Çırpan, "Spatial modulation GFDM: A low complexity MIMO-GFDM system for 5G wireless networks," in *IEEE International Black Sea Conference on Communications and Networking (BlackSeaCom)*, Jun. 2016, pp. 1–5.

[20] ——, "Generalized frequency division multiplexing with space and frequency index modulation," in *IEEE International Black Sea Conference on Communications and Networking*, Dec. 2017.